\documentstyle{mn}

\newcommand{\magper}{mag~arcsec$^{-2}$}
\newcommand{\micr}{\,$\mu$m}
\newcommand{\bd}{B_d}
\newcommand{\bm}{B_m}
\newcommand{\dpolm}{P_m}

\input{epsf}

\title[Optimising background-limited observing]
{Optimising background-limited observing \\
during bright-moon phases and twilight}

\author[I.~K.~Baldry and J.~Bland-Hawthorn]
{I.~K.~Baldry and J.~Bland-Hawthorn\\
Anglo-Australian Observatory, P.O.\,Box 296, Epping, NSW 1710, Australia}

\date{Preprint. Accepted 2000 October}

\pagerange{\pageref{firstpage}--\pageref{lastpage}}

\pubyear{2000}

\begin{document}

\maketitle
\label{firstpage}

\begin{abstract}
  For the majority of optical observing programs, the sky brightness
  provides the fundamental limit to signal detection such that the
  scientific feasibility is largely dictated by the moon phase.  Since
  most observatories do not have the resources to build expensive
  high-resolution or infrared instruments, they are increasingly at a
  loss as to how to exploit bright time. We show that, with due
  consideration of the field and moon position, it is possible to
  undertake `dark time' observing programs under `bright time'
  conditions. Our recommendations are particularly appropriate to
  all-sky survey programs.

  In certain instances, there are gains in observing efficiency with
  the use of a polariser, which can significantly reduce the moonlight
  (or twilight) sky-background flux relative to an extraterrestrial
  flux. These gains are possible in background-limited cases because
  the sky background can be highly polarised, due to scattering,
  around ninety degrees away from the moon (or sun). To take advantage
  of this, only minor modifications to existing instruments are
  needed.
\end{abstract}

\begin{keywords}
  instrumentation: miscellaneous --- methods: observational
\end{keywords}

\section{Introduction}

To make effective use of bright time, ground-based optical telescopes
are generally used to observe either sufficiently bright objects in
the optical ($V \la 18$ \magper) or to observe in the near infrared.
In the first case, this is to avoid being background limited by the
moonlit sky, and in the second case, the sky background is dominated
by OH line emission\footnote{Note that if the spectral resolution is
  high enough to observe between the OH lines in the 1--2.2\micr\ 
  wavelength range, then background continuum becomes important and
  dark time conditions may then be preferable.} or thermal emission
from the sky and telescope.  For many telescopes, the observatory
budget does not extend to expensive high-resolution optical
spectrographs or infrared instruments which are suitable for
bright-time observations.  Some of these telescopes have formed
consortia which span the globe to undertake monitoring of variable
sources (e.g., pulsating stars, microlensing events, asteroids). 

In this paper, we look at strategies for observing during bright time
that allow background-limited objects to be observed with higher
efficiency.  We also look at the possibility of improving observations
during twilight.  In the optical, the moonlit sky is generally darker
along sight lines at large angles from the moon.  Additionally, when
the moon is near ninety degrees away, the scattered moonlight is
highly polarised.  By using a polarising filter oriented at the
correct angle, the sky background can then be reduced to near-dark or
grey conditions.

The degree of polarisation can be defined as
\begin{equation}
  P = \frac{I_p}{I_p + I_u} \:\: {\rm or} \:\:
  \frac{I_1 - I_2}{I_1 + I_2} \: ,
\end{equation}
where: $I_p$ is the polarised intensity and $I_u$ is the unpolarised
intensity, or; $I_1$ is the maximum intensity as measured in one plane
of polarisation and $I_2$ is the intensity in the orthogonal plane.
Studies of the polarisation of twilight (e.g., Coulson 1980, Wu \& Lu
1988) have shown that scattered sunlight at ninety degrees has a
degree of polarisation of about 0.85. Moonlight scattered at ninety
degrees could have a slightly higher degree of polarisation since the
light is partially polarised on reflection from the moon (e.g.,
Dollfus 1962).

The next section quantifies observing efficiency for Poisson-noise
limited observations, and determines the gain factor when using a
polariser. Sections~\ref{sec:obs-moonlit} and~\ref{sec:obs-twilight} 
discuss observations in moonlight and twilight with and without a 
polariser.

\section{Observing efficiency}
\label{sec:obs-eff}

A figure of merit (proportional to observing efficiency), 
for a particular instrument observing an object of
a certain flux, can be defined as the signal-to-noise ratio squared
($R^2$) divided by the integration time ($t$).  Ignoring read-out
noise, this is given by\footnote{Equations~\ref{eqn:snr-no-pol}
  and~\ref{eqn:snr-pol} relate to observations where the background
  can be estimated over many more pixels than the object, otherwise
  the background is effectively higher (e.g., a factor of two in the
  case of beam switching with fibres).}
\begin{equation}
  f = \frac{R^2}{t} \simeq \frac{\epsilon O^2}{O + \bd + \bm} \: ,
  \label{eqn:snr-no-pol}
\end{equation}
where $\epsilon$ is the efficiency of the telescope and instrument,
$O$ is the object flux on the primary mirror,
$\bd$ is the background flux due to the {\em dark} sky
(could also include dark current) and
$\bm$ is the extra background flux due to
{\em moonlight} and/or twilight (i.e., $\bm=0$ for dark-time observations).

Assuming the object and dark-sky fluxes are unpolarised and a polariser
is placed in the beam to minimise the extra background flux, 
the figure of merit is then given by
\begin{equation}
  f' \simeq \frac{\tau \epsilon' O^2}
  {O + \bd + (1 - \dpolm) \bm + 2 \gamma \dpolm \bm} \: ,
  \label{eqn:snr-pol}
\end{equation}
where $\tau$ is the transmission of the polariser with unpolarised light
($\tau \le 0.50$),
$\epsilon'$ is the efficiency of the telescope and instrument
with polarised light,
$\gamma$ is the extinction\footnote{We define the extinction of a
  polariser as the ratio of the minimum and maximum transmissions of
  polarised light, obtained by rotation.} of the polariser ($\gamma \ll 1$)
and $\dpolm$ is the degree of polarisation of the extra light.

The gain in observing efficiency from using a polariser can be written as
\begin{equation}
  \frac{f'}{f} \simeq \tau \frac{\epsilon'}{\epsilon} \left[ 
  \frac{O + \bd + \bm}
  {O + \bd + (1 - \dpolm  + 2 \gamma \dpolm) \bm} \right] \: .
  \label{eqn:gain}
\end{equation}
There may also be an extra gain factor due to an improved duty cycle
because the background level is lower and, therefore, integration
times can be increased.  In most cases, $\epsilon' / \epsilon$ will be
approximately unity\footnote{Some existing spectrographs will have a
  significantly higher throughput in one polarisation state. We also
  note the possibility of designing a spectrograph to be optimised for
  one polarisation. For example, this is possible using volume phase
  holographic gratings.} and $\tau$ will be in the range 0.35--0.48
(Sec.~\ref{sec:polarisers}).  The degree of polarisation could be
about 0.90 for moonlight scattered from the sky ninety degrees away
from the moon. The extinction should not be a limiting factor with
$\gamma$ in the range $10^{-3}$--$10^{-2}$.

Setting $\gamma = 0$, $\epsilon' / \epsilon = 1$ and rearranging
Equation~\ref{eqn:gain}, we obtain a condition for gains of greater
than or approximately unity ($f'/f \ga 1$):
\begin{equation}
  (\dpolm + \tau - 1) \bm \ga (1 - \tau) (O + \bd) \: .
  \label{eqn:for-gain}
\end{equation}
This condition can also be written as
\begin{equation}
  (P + \tau - 1) B \ga (1 - \tau) O \: ,
  \label{eqn:for-gain2}
\end{equation}
where $P$ is the degree of polarisation of the total background flux
($B$).  In other words, $P > 1 - \tau$ and the background flux must
be high enough for a polariser to improve observing efficiency.  For
example, with $P = 0.78$ and $\tau = 0.48$, the background flux
must be greater than or approximately twice the object flux.

\subsection{Polarising filters}
\label{sec:polarisers}

Normal polarising filters are notoriously lossy with typical
transmissions of 0.70 (one polarisation state).  However, thin film
polarisers made by `bleaching' can have transmissions as high as 0.96
(Gunning \& Foschaar 1983). Here, a polymerising sheet is placed in
acetone to eat away most of the plastic.  The polarising film is then
stretched and deposited over a substrate in a humid oven. Much of the
art of bleaching resides in technical reports of the Lockheed Palar
Alto Research Laboratory where it was first developed by H.~E.~Ramsey.
When achieving the highest transmissions through bleaching, it is
important that the extinction ratio be kept to about 10$^{-2}$ or
lower.

\section{Observing moonlit sky}
\label{sec:obs-moonlit}

\begin{figure*}
\epsfbox{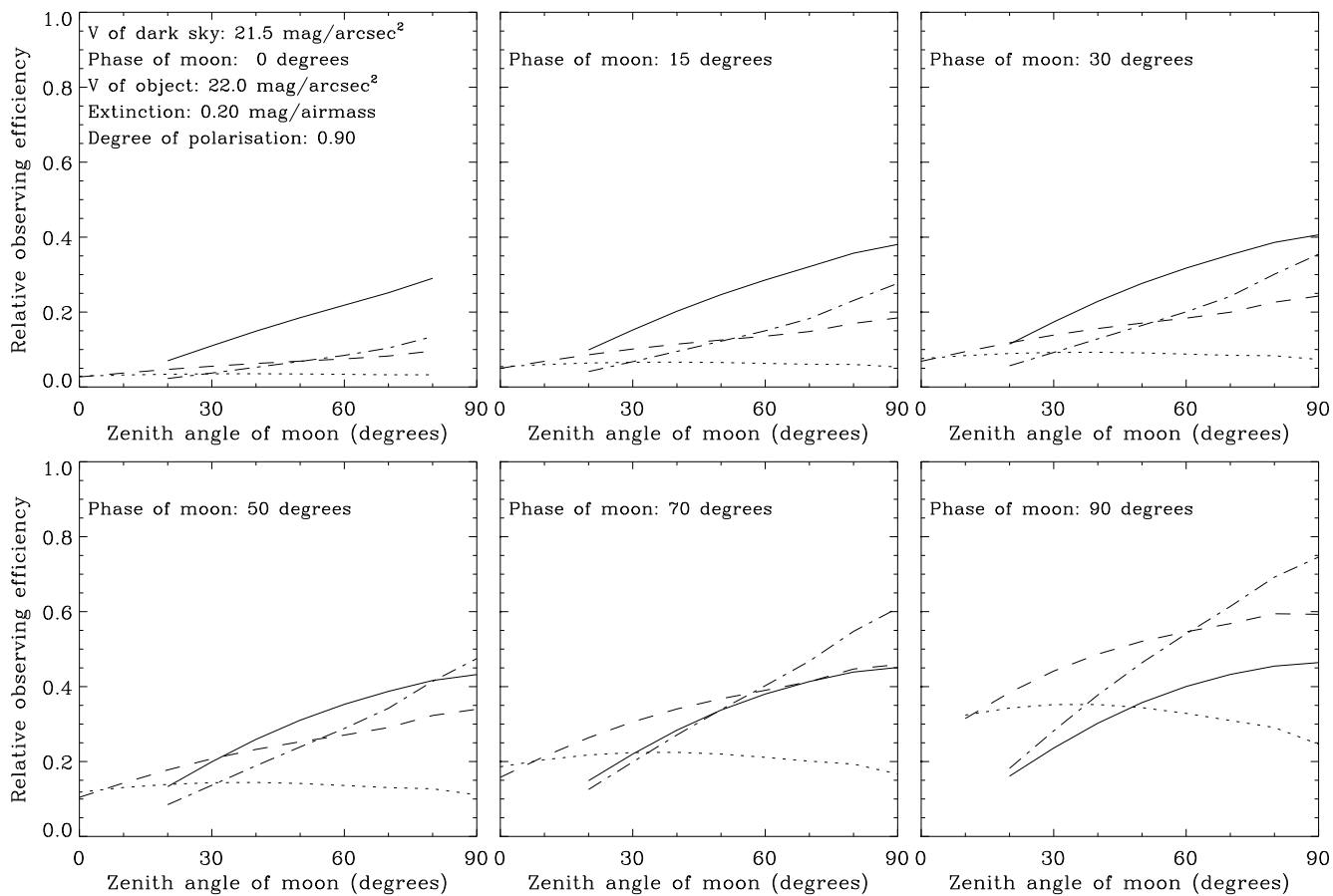}
\caption{Calculated observing efficiencies at various moon phases ($\alpha$)
  from full moon to half moon, relative to dark-time observations at
  zenith. The dotted line represents observations taken 30\degr\ away
  from the moon, the dashed line 60\degr\ away and the dash-and-dotted
  line 90\degr\ away.  The solid line represents observations taken
  90\degr\ away with a polariser used to minimise the sky background,
  assuming the degree of polarisation of the moonlight ($\dpolm$) is
  0.90 and the transmission of the polariser ($\tau$) is 0.48 with
  unpolarised light.  The efficiencies are calculated using a model of
  the brightness of moonlight, described by Krisciunas \& Schaefer
  (1991), and taking into account the extinction of object flux at the
  lowest possible zenith angle (for each moon zenith angle and
  moon-object separation angle).}
\label{fig:eff-moon}
\end{figure*}

To consider the effect of observing objects during moonlit sky, we
used the model of moonlight brightness described by Krisciunas \&
Schaefer (1991).  For various moon phases ($\alpha$)\footnote{The phase
  $\alpha$ is defined as the angular distance between the Earth and
  the Sun as seen from the Moon.}, zenith angles and moon-object
separation angles, we calculated the observing efficiency (using
Equations~\ref{eqn:snr-no-pol} and~\ref{eqn:snr-pol}) relative to
dark-time observations at zenith.  The extinction of object flux was
considered in addition to the variation in background flux.
Figure~\ref{fig:eff-moon} shows the results of these calculations for
observations of $V=22$ \magper\ objects during phases from full moon
to half moon.

During moonlit bright time ($|\alpha| < 60\degr$), it is generally
more efficient to observe {\em with a polariser} than without (on
appropriate targets, i.e., background limited and near 90\degr\
separation from the moon).  Remembering that Poisson-noise gains of
less than but near unity (Eqn.~\ref{eqn:gain}) can provide better
observing efficiency if the duty cycle is significantly improved.
Additionally, residuals associated with continuum-sky subtraction may
be reduced since the relative contribution of the moonlight (solar
spectrum) to the overall flux is lower.

During moonlit grey time ($60\degr < |\alpha| < 120\degr$), it is
generally best to observe without a polariser on targets
that are about 40\degr\ to 90\degr\ away from the moon.

\begin{figure*}
\epsfbox{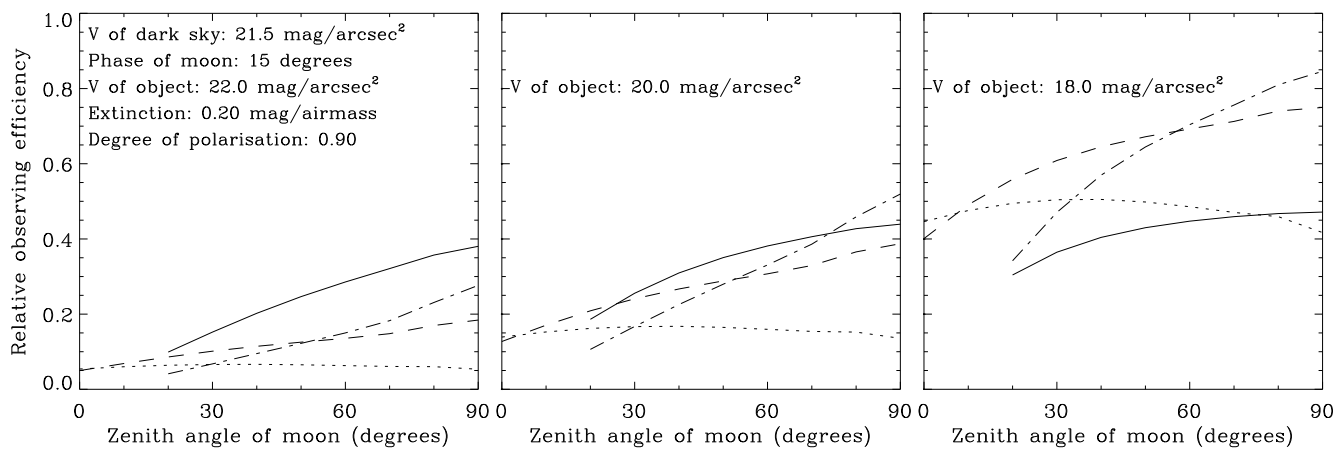}
\caption{Calculated observing efficiencies near full moon
  for objects of 22, 20 and 18 \magper, relative to dark-time
  observations at zenith. See Figure~\ref{fig:eff-moon} for details.}
\label{fig:eff-object}
\end{figure*}

Figure~\ref{fig:eff-object} shows the observing efficiency for various
object brightnesses about one day away from full moon (relative to
dark-time observations of the same object brightness).  There is some
improvement when using a polariser for objects fainter than about 20
\magper\ at this moon phase. The brightness for which it is useful to
use a polariser depends on the moonlight, the degree of polarisation,
the dark-sky contribution, transmission and instrument efficiencies
(see Equation~\ref{eqn:for-gain}).

\section{Observing twilight sky}
\label{sec:obs-twilight}

\begin{figure*}
\epsfxsize=14.0cm
\centerline{
\epsfbox{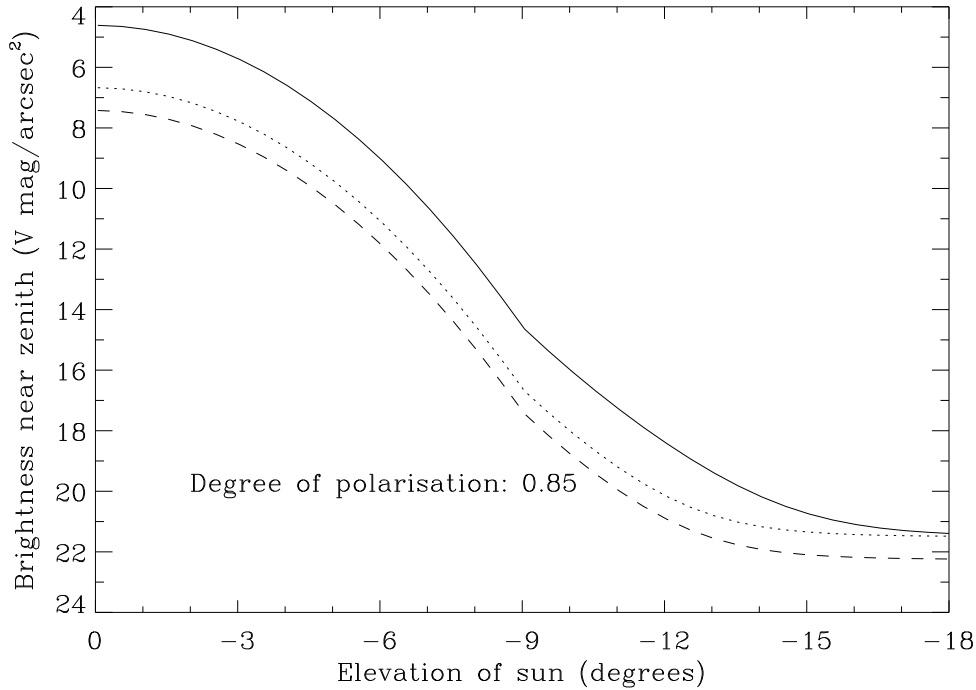}
}
\caption{Model brightness of the sky during twilight (with no moon).
  The solid line represents the total flux, while the dashed line
  represents the flux of the minimum-intensity plane of polarisation,
  assuming the the degree of polarisation of the twilight is 0.85 and
  the dark-sky contribution is unpolarised. The dotted line is the
  same, magnitude adjusted by a factor of two in flux.  The dark-sky
  brightness was taken as 21.5 \magper\ and the twilight brightness
  was similar to the model described by Tyson \& Gal (1993).}
\label{fig:degrees-bright}
\end{figure*}

To consider observations during twilight, we produced a model of the
twilight brightness based on the work by Tyson \& Gal (1993). The sky
brightness versus the elevation angle of the Sun is shown in
Figure~\ref{fig:degrees-bright}.  At sunset, the brightness was
assumed to be about 4.5 \magper\ decreasing towards the dark-sky
brightness at the end of astronomical twilight, where we assumed that
the twilight contribution was 0.1 times the dark-sky contribution. The
flux for the minimum-intensity plane of polarisation, observing about
90\degr\ from the Sun, is also shown.

\begin{figure*}
\epsfbox{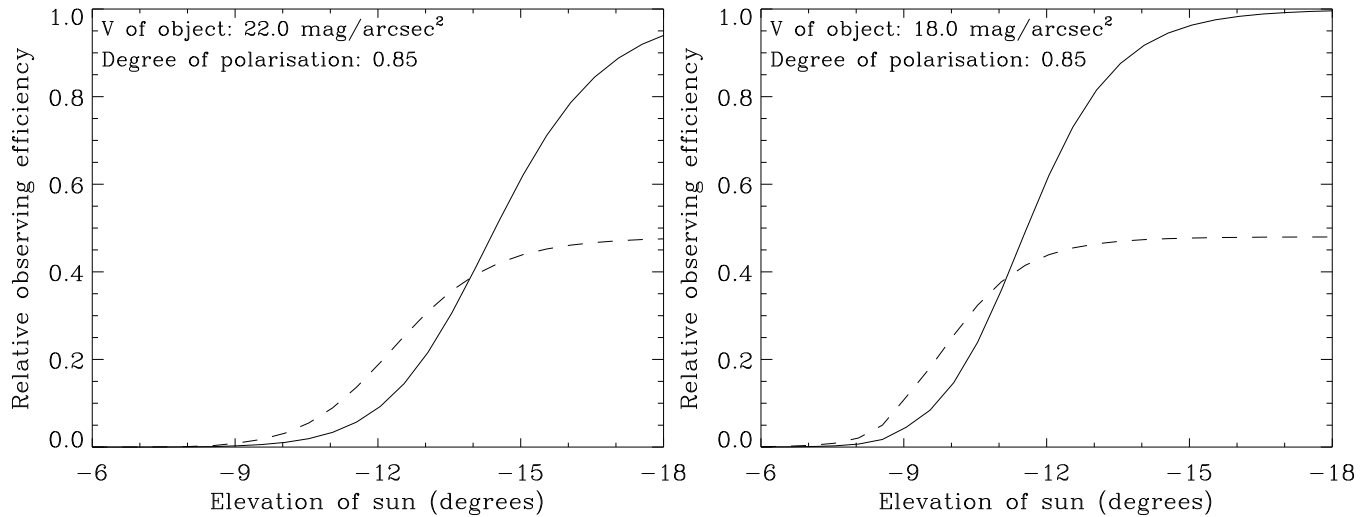}
\caption{Calculated observing efficiencies during twilight
  for objects of 22 and 18 \magper, relative to dark-time observations
  at zenith.  The solid line represents normal observations near
  zenith, while the dashed line represents observations taken with a
  polariser used to minimise the sky background
  (Fig.~\ref{fig:degrees-bright}).}
\label{fig:degrees-eff-obs}
\end{figure*}

Figure~\ref{fig:degrees-eff-obs} shows the Poisson-noise observing
efficiencies for two different brightness objects during twilight.
There is only a small range in elevation ($\sim 2 \degr$) where the
efficiency is above 0.2 with a polariser and where the gain is about
or above unity (Eqn.~\ref{eqn:gain}). Therefore, there is probably
little use for this type of technique during twilight, with the
possible exception of observing at high latitude sites, e.g., on the
Antarctic continent where polarisers could allow observing in much
brighter conditions which can last days or weeks.  There, they could
be used for site testing even if there are no specific science cases
for twilight observing.  By extension, a polariser could be used for
daytime observations (at any latitude) such as astrometric star or
planet measurements (e.g., Rafferty \& Loader 1993).

\section{Discussion}

For effective use of a polariser to minimise the sky background, it is
necessary to be able to rotate the polariser or the instrument in a
short time relative to exposure times.  A minor adaption to
instruments that are in operation or under design is required,
including software to calculate the required rotation angle from the
moon and object celestial coordinates.  To check the validity of the
setup, the sky flux can be measured at various rotation angles (of the
polariser or instrument) in order to determine the angle that
transmits the minimum sky flux.

For observatories that aim to optimise the use of bright-grey
time for background limited observations, observing proposals and
telescope schedules need to consider the moon's position.

Potentially, the most efficient use of bright-time observing with or
without a polariser will be for those instruments where
$\epsilon'/\epsilon$ can be significantly greater than unity.
Equation~\ref{eqn:gain} refers to the gain relative to the average of
$f$ over instrument rotation angles. The ability to rotate the
instrument independently of the polariser gives the option to maximise
$\epsilon'/\epsilon$ or, gains can be made without a polariser just by
rotating the instrument so that the background flux is minimised.  The
factor $\epsilon'/\epsilon$ could be measured for existing
spectrographs, and in the future, spectrographs could be optimised for
one polarisation state.

Under design at the AAO is a tunable Lyot filter that is phase free
over a wide field (Bland-Hawthorn et al.\ 2000). The Lyot
filter requires a single polarisation input and therefore such an
instrument has effectively $\epsilon'/\epsilon = 2$ at the appropriate
polarisation angle, i.e., a polariser is used in any case. Such an
instrument could be used during bright-grey time with instrument
rotation used to minimise the sky background without loss of object
flux.

\subsection{Further work}

To determine more accurately the gain in using a polariser, it is
necessary for more measurements to be made.  These include
measurements of the polarisation of the moonlight contribution to the
sky background, as a function of separation angle from the moon and as
a function of wavelength. This is not straightforward because the
moonlight contribution can not easily be distinguished from the
dark-sky contribution. Important questions to answer are: over what
range of the sky is it useful to use a polariser, e.g., 80\degr\ to
100\degr\ from the moon or only 85\degr\ to 95\degr; for which
broad-band filters is it useful, e.g., $BVRIz$ or just $V$; in the
$RIzJH$ bands, how is narrower-band imaging or spectroscopy between
the OH emission lines affected; how does the degree of polarisation
vary with atmospheric conditions?

\section{Conclusions}

We suggest there is a key role to play for telescopes that do not have
traditional bright-time instruments.  Near all-year-round dark-grey
conditions can be obtained for survey work (albeit sometimes at lower
efficiency), by consideration of the moon's position and the use of a
polariser during bright time.  For any telescope, better use can be
made of bright-grey conditions with or without a polariser.

\section*{Acknowledgements}

IKB would like to thank Tim Bedding and Hans Kjeldsen, and he hopes
that this paper does not result in reduced observing time for stellar
oscillation studies.

\bsp
\label{lastpage}

\end{document}